\documentclass[9pt,twocolumn,twoside]{pnas-new}

\templatetype{pnasresearcharticle} 

\begin{document}

\title{Maximum spreading of impacting shear-thinning and shear-thickening drops}

\author[a]{Anahita Mobaseri}
\author[a]{Satish Kumar}
\author[a,b,1]{Xiang Cheng}

\affil[a]{Department of Chemical Engineering and Materials Science, University of Minnesota, Minneapolis, Minnesota 55455, USA}
\affil[b]{St. Anthony Falls Laboratory, University of Minnesota, Minneapolis, Minnesota 55414, USA}

\leadauthor{Mobaseri}

\significancestatement{How much does a liquid drop spread upon impacting a solid surface? The apparent simplicity of this question, coupled with the ubiquity of drop impact in daily life, belies the complexity of the underlying fluid dynamics and its broad relevance across applications, from printing and coating to forensics. While the maximum spreading of Newtonian drops has been well studied, predicting that of non-Newtonian drops remains challenging due to significant viscosity variations during impact. By integrating simulations, theory, and experiments, we connect the impact of Newtonian and non-Newtonian drops and quantify the maximum spreading of the latter. This work not only advances the basic understanding of drop impact but also offers practical insights for optimizing drop-impact-driven processes through tailored fluid properties.}

\authorcontributions{Conceptualization: A.M, S.K., and X.C. Methodology: A.M, S.K., and X.C. Simulation, experiment, and data analysis: A.M. Project administration: X.C. Supervision: S.K, X.C. Writing—original draft: A.M. Writing—review and editing: S.K., X.C.}
\authordeclaration{The authors declare no competing interest.}
\correspondingauthor{\textsuperscript{1}To whom correspondence should be addressed. E-mail: xcheng@umn.edu}


\begin{abstract}
The maximum spreading of an impacting liquid drop is a key metric for characterizing the fundamental fluid process of drop impact. While extensively studied for Newtonian liquids, how far a non-Newtonian drop can spread upon impacting a solid substrate remains an open question. Here, by combining simulations, experiments, and scaling analyses, we establish a general framework for predicting the maximum spreading of drops of generalized Newtonian liquids, encompassing both shear-thinning and shear-thickening behaviors. Through an analysis of the energy budget at maximum spreading, we identify a characteristic shear rate that governs the viscous dissipation during drop impact. The finding allows us to map the spreading of non-Newtonian drops onto that of Newtonian drops, revealing the quantitative dependence of the maximum spreading diameter on various impact parameters and rheological properties of liquids. Our study addresses the long-standing challenge of understanding the impact dynamics of non-Newtonian drops, and provides valuable guidance for designing non-Newtonian liquids to achieve desired impact outcomes.
\end{abstract}

\dates{This manuscript was compiled on \today}
\doi{\url{www.pnas.org/cgi/doi/10.1073/pnas.XXXXXXXXXX}}

\maketitle
\thispagestyle{firststyle}
\ifthenelse{\boolean{shortarticle}}{\ifthenelse{\boolean{singlecolumn}}{\abscontentformatted}{\abscontent}}{}

\firstpage[1]{4}

\dropcap{T}he impact of liquid drops on solid surfaces is a ubiquitous fluid phenomenon relevant to many natural and industrial processes and has been extensively studied for over a century, dating back to the pioneering work of Worthington \cite{Yarin2006Review,Josserand2016Review,Cheng2022Review}. Among all the quantities characterizing drop impact, the maximum spreading of impacting drops garners the most research interest due to its direct influence on the outcome of a drop impact event. Accurately predicting the maximum spreading of an impacting drop is essential for understanding natural phenomena, such as the size of raindrop craters on soil \cite{zhao2015granular,zhang2015granular}, and for optimizing industrial processes, ensuring the quality of coated layers in spray coating \cite{ye2017analysis, wardhono2023fluid} and inkjet printing \cite{derby2003inkjet,guo2024spreading}, as well as improving anti-fouling and drying efficiency in food processing \cite{balzan2021drop, duan2022study}. The maximum spreading of blood drops is also a critical parameter in forensic science for reconstructing events at crime scenes \cite{hulse2005deducing}.

Previous studies have established various relations for the maximum spreading diameter $D_{\text{max}}$ of Newtonian drops based on impact parameters in different impact regimes (see a summary in Ref.~\cite{Aksoy2022Spreaing}). 
These impact parameters are commonly represented by two dimensionless numbers: the Reynolds number $Re = \rho U_0 D_0 / \mu$ that compares inertial and viscous forces, and the Weber number $We = \rho U_0^2 D_0 / \sigma$ that compares inertial and capillary forces. Here, $U_0$ and $D_0$ are the impact velocity and diameter of the liquid drop and $\rho$, $\mu$, and $\sigma$ are the density, viscosity, and surface tension of the liquid. In the viscous regime, the maximum spreading diameter, $D_{\text{max}}$, is estimated by balancing kinetic energy and viscous dissipation, which leads to a scaling relation $D_{\text{max}}/D_0 \sim Re^{1/5}$ \cite{chandra1991collision,clanet2004maximal}. In the capillary regime with negligible viscous dissipation, the balance between kinetic and surface energy yields $D_{\text{max}}/D_0 \sim We^{1/2}$ \cite{bennett1993splat}. To bridge the viscous and capillary regimes, Eggers et al. introduced an impact number $P = We Re^{-2/5}$, where $P\gg 1$ corresponds to the viscous regime and $P\ll1$ to the capillary regime \cite{eggers2010drop}. Laan and co-workers further applied a Padé approximant to fit the dependence of $D_{\text{max}}$ on $P$, which provides a good description of experimental results \cite{laan2014maximum}.

While extensive studies have been conducted on the maximum spreading of Newtonian drops, our understanding of the maximum spreading of non-Newtonian drops remains incomplete \cite{Shah2024_review, liu2024universal, an2012maximum, balzan2021drop, dechelette2010non, yokoyama2022droplet, german2009impact,isukwem2024role,isukwem2025viscoplastic,luu2009drop,blackwell2015sticking,oishi2019normal}, despite their widespread industrial applications and prevalence in biological systems. In particular, for generalized Newtonian liquids that exhibit strong shear thinning or shear thickening, the challenge arises from the substantial variation in shear rate and the resulting highly non-uniform viscosity distribution within an impacting drop. The complexity renders most theoretical approaches---typically based on the assumption of constant viscosity---ineffective.

The most common approach to address the complexity of the impact of a drop of a generalized Newtonian liquid is to assume a characteristic shear rate of drop impact, $\dot{\gamma}_c$, or equivalently a characteristic length of drop impact $L_c$ via $\dot{\gamma}_c \equiv U_0/L_c$. The effective viscosity, $\mu_c = \mu(\dot{\gamma}_c) = \mu(U_0/L_c)$, is then estimated using the non-Newtonian constitutive relation of the liquid. A mapping is finally performed by relating the maximum spreading of the non-Newtonian drop to the well-studied maximum spreading of a Newtonian drop with constant viscosity $\mu_c$ \cite{laan2014maximum,de2021droplet,boyer2016drop,shah2022coexistence,luu2009drop, isukwem2024role, liu2024universal, an2012maximum,quirke2024spreading}. Although the very existence of $\dot{\gamma}_c$ enabling the mapping between non-Newtonian and Newtonian drop impact remains unverified, three different models have already been proposed for the choice of $\dot{\gamma}_c$: (\textit{i}) The shear rate within an impacting drop is so large that the infinite-shear viscosity is sufficient to describe non-Newtonian drop impact \cite{laan2014maximum, de2021droplet}. Thus, this model essentially assumes $\dot{\gamma}_c \to \infty$ and $L_c \to 0$. (\textit{ii}) The characteristic shear rate is determined by the velocity and length scale of the impacting drop, i.e., $\dot{\gamma}_c = U_0/D_0$ with $L_c = D_0$ \cite{boyer2016drop, shah2022coexistence}. (\textit{iii}) The strongest shear, occurring at the smallest length scale of the drop, dominates the impact process. Thus, $\dot{\gamma}_c = U_0/H_{\text{min}}$ \cite{luu2009drop, liu2024universal, an2012maximum, quirke2024spreading, isukwem2024role}, where $L_c = H_\text{min}$ is the smallest height of the drop, which is achieved when the drop reaches its maximum spreading. This height is further approximated as $H_\text{min} \approx 2D_0^3/(3D_\text{max}^2)$, assuming a puddle-shaped drop and conservation of volume.

\begin{figure*}[t]
    \includegraphics[width=1\linewidth]{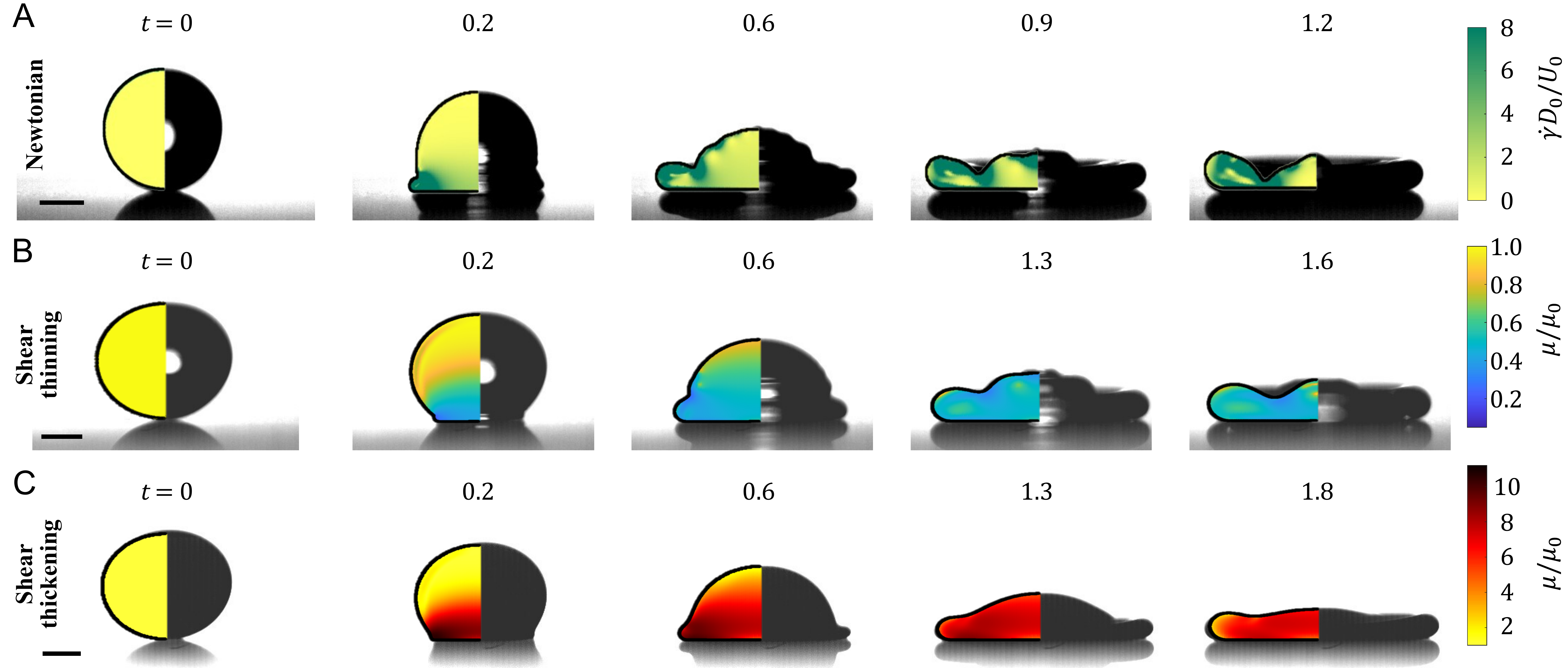}
    \caption{Impact dynamics of Newtonian, shear-thinning and shear-thickening drops. The time sequences compare our simulations (left) with corresponding experiments (right). The time of each snapshot, normalized by $D_0/U_0$, is shown above, where $D_0$ is the drop diameter and $U_0$ is the impact velocity. (\textit{A}) Impact of a water drop with $U_0 = 0.8$ m/s, $D_0 = 2.8$ mm, viscosity $\mu = 1$ mPa$\cdot$s, and surface tension $\sigma = 72$ mN/m. The Reynolds number and the Weber number of the impact are $Re = 2240$ and $We = 25$, respectively. The color in the simulation indicates shear rate $\dot{\gamma}$, normalized by $U_0/D_0$. (\textit{B}) Impact of a shear-thinning drop made of a 0.03 wt\% xanthan gum aqueous solution with $U_0 = 0.5$ m/s, $D_0 = 3.4$ mm, and $\sigma = 72$ mN/m. The rheological properties of the solution are measured experimentally and fit to the Carreau model (Eq.~\ref{eq:Carreau_model})(SI Sec.~C), yielding $\mu_0 = 17$ mPa$\cdot$s, $\beta = 0.05$, $n=0.71$ and $\lambda = 3.90 D_0/U_0 = 0.012$ s. The zero-shear Reynolds number $Re_0 = 102$ and $We = 12$. (\textit{C}) Impact of a shear-thickening drop made of a 36 wt\% cornstarch-water mixture with $U_0 = 0.7$ m/s, $D_0 = 3.4$ mm, $\sigma = 72$ mN/m, $\mu_0 = 5.5$ mPa$\cdot$s, $\beta = 11.20$, $n=0.1$, and $\lambda = 0.85 D_0/U_0 = 0.004$ s. $Re_0 = 455$ and $We = 27$. The color in the simulation of (\textit{B}) and (\textit{C}) shows the normalized viscosity, $\mu/\mu_0$. The contact angle of simulations is $\theta = 180^\circ$, whereas $\theta = 150 \pm 5^\circ$ in experiments. The scale bars represent 1 mm. See also Movies 1 and 2 for shear-thinning and shear-thickening drop impact from experiments.}
    \label{fig:Figure 1}
\end{figure*}

To validate the hypothesis of a characteristic shear rate and resolve the ongoing debate over its correct formulation, we conduct numerical simulations on the drop impact of generalized Newtonian liquids exhibiting shear thinning and shear thickening behaviors. Through a systematic analysis of energy conversion among kinetic energy, surface energy, and viscous dissipation during drop impact, we demonstrate the existence of a characteristic shear rate for describing the maximum spreading of drop impact and uncover its dependence on various impact parameters, which challenges all three existing models. The finding is further supported by a scaling analysis on the asymptotic film thickness determined by the interplay between the free surface of the falling drop and the growing boundary layer within the drop. Leveraging this finding, we map the impact dynamics of non-Newtonian drops onto that of Newtonian drops and derive a formula for the maximum spreading diameter of non-Newtonian drops. The prediction quantitatively matches both existing experimental data from the literature and our own measurements on various non-Newtonian liquids. Taken together, our study addresses one of the key open questions in the study of drop impact and paves the way to control drop impact dynamics by engineering the rheological properties of liquids.

\section*{Results}

\begin{figure}[ht]
    \centering
    \includegraphics[width=0.9\linewidth]{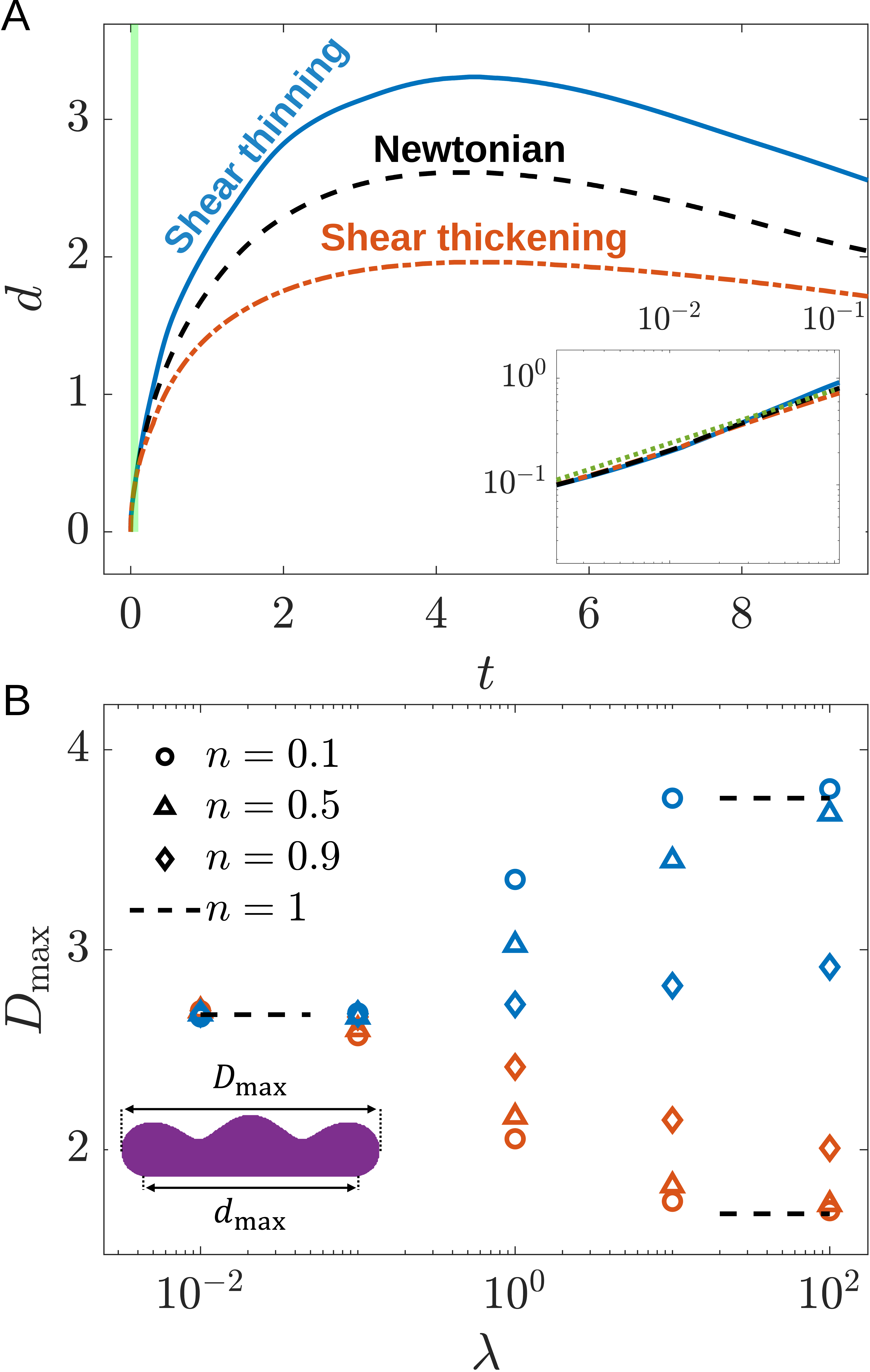}
    \caption{Spreading of impacting drops from simulations. (\textit{A}) Temporal evolution of the diameter of the contact area $d(t)$ during the spreading of Newtonian, shear-thinning, and shear-thickening drops with $Re_0 = 200$ and $We = 556$. For non-Newtonian liquids, the power-law index $n = 0.1$ and the time constant $\lambda = 1$. Viscosity ratio $\beta = 0.1$ for the shear thinning liquid and $\beta = 10$ for the shear-thickening liquid. The green region marks the early-impact regime shown in the inset. Inset: Early-time dynamics of $d(t)$. The green dotted line indicates $d=\sqrt{6t}$. (\textit{B}) Maximum spreading diameter $D_{\text{max}}$ of shear-thinning (blue symbols) and shear-thickening (red symbols) drops versus $\lambda$ at different $n$. The dashed line on the left indicates $D_{\text{max}}$ of a Newtonian drop of viscosity $\mu_0$, while the dashed lines on the right indicate $D_{\text{max}}$ of Newtonian drops of viscosity $\mu_\infty$ with $\mu_\infty = 0.1\mu_0$ (top) and $10\mu_0$ (bottom), respectively. Inset: A schematic showing the definitions of the diameter of the maximum contact area $d_{\text{max}}$ and the maximum spreading diameter $D_{\text{max}}$.}
    \label{fig:Figure 2}
\end{figure}

\subsection*{Impact dynamics of non-Newtonian drops}
We perform direct numerical simulations of drop impact based on the volume-of-liquid solver Basilisk, which solves the time-dependent mass and momentum conservation equations with second-order accuracy in both space and time \cite{basilisk2013}. To capture the shear-rate-dependent viscosity of non-Newtonian liquids, we employ the Carreau model, a constitutive model for generalized Newtonian liquids \cite{bird1976useful,carreau1979analysis,Macosko1994}:
\begin{equation}
    \frac{\mu(\dot{\gamma})}{\mu_0} = \beta + (1 - \beta) \left[ 1 + (\lambda \dot{\gamma})^2 \right]^{\frac{n-1}{2}}, \label{eq:Carreau_model}
\end{equation}
where $\beta = \mu_{\infty}/\mu_0$ represents the ratio of the infinite-shear viscosity $\mu_{\infty}$ to the zero-shear viscosity $\mu_0$, $\lambda$ is the time constant that governs the onset of shear-thinning or shear-thickening behavior, and $n$ is the power-law index controlling the sharpness of the transition between the zero-shear and the infinite-shear viscosity. The shear rate is $\dot{\gamma} = \sqrt{2 \Pi_D}$, where $\Pi_D$ is the second invariant of the rate-of-deformation tensor of flow velocity $\bf{u}$, $\bf{D} = \nabla \bf{u} + (\nabla \bf{u})^T$ \cite{Macosko1994}. Shear thinning is represented by $\beta <1$, while shear thickening corresponds to $\beta > 1$. Newtonian behavior is recovered when $\beta = 1$ (or $n = 1$). To focus on the influence of non-Newtonian rheology on viscous dissipation, we simulate drop impact on a superhydrophobic surface with a fixed contact angle $\theta = 180^\circ$. The effect of surface wettability and contact angle on the maximum spreading will be discussed at the end when comparing our numerical predictions with experiments. Further details about the Carreau model and our numerical method can be found in Supporting Information (SI) Sec.~A.

The impact dynamics of Newtonian, shear-thinning and shear-thickening drops from our simulations show excellent agreement with experiments at the same impact and rheological parameters (Fig.~\ref{fig:Figure 1}). The detailed flow field from our simulations reveals the rapidly evolving non-uniform shear rate in the lower region of the drop upon impact (Fig.~\ref{fig:Figure 1}\textit{A}). Near the contact line at the base of the spreading lamella, high shear rate emerges due to the steep velocity gradient that redirects the flow from the vertical to the radial direction \cite{eggers2010drop, sun2022stress, liu2024universal,Philippi2016_dropimpact,Gordillo2018_impactforce}. The heterogeneous shear rate leads to substantial variations of viscosity inside the non-Newtonian drops (Figs.~\ref{fig:Figure 1}\textit{B} and \textit{C}), which profoundly affect the maximum spreading of the drops. 

\begin{figure*}[t]
    \includegraphics[width=1\linewidth]{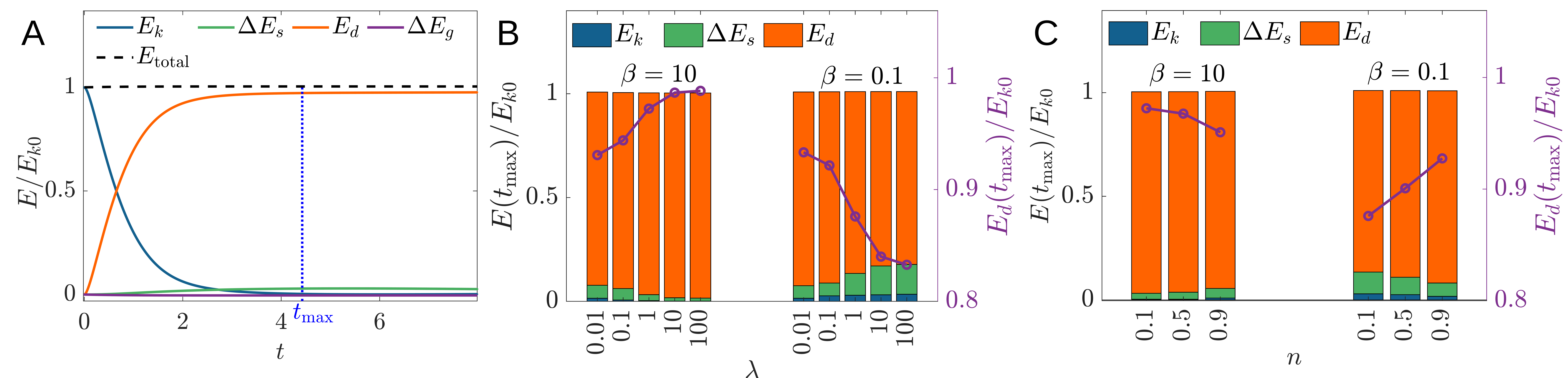}
    \caption{Energy budget of impacting non-Newtonian drops. $Re_0 = 200$ and $We = 556$. (\textit{A}) Energy conversion among kinetic energy $E_k$, the change of surface energy $\Delta E_s$, viscous dissipation $E_d$, and the change of gravitational energy $\Delta E_g$ during the impact of a shear-thickening drop with $\beta=10$, $\lambda=1$, and $n=0.1$. The summation of all energies is constant and equal to the initial kinetic energy $E_{k0}$, as dictated by energy conservation. The time of maximum spreading, $t_{\text{max}}$, is marked by a vertical dotted line. Energy distribution at the maximum spreading for shear-thickening ($\beta = 10$) and shear-thinning ($\beta = 0.1$) drops at different $\lambda$ (\textit{B}) and $n$ (\textit{C}). Empty circles in (\textit{B}) and (\textit{C}) (right axes) denote the fraction of the viscous dissipation, $E_d(t_\text{max})/E_{k0}$, at the maximum spreading. $\Delta E_g$ remains negligible throughout the impact and is therefore omitted in (\textit{B}) and (\textit{C}) for clarity. Nevertheless, $\Delta E_g$ is accounted for in all our calculations.}
    \label{fig:Figure 3}
\end{figure*}

Hereafter, we non-dimensionalize all the quantities by using the length scale $D_0$, the time scale $D_0/U_0$, and the mass scale $\rho D_0^3$. Figure~\ref{fig:Figure 2}\textit{A} shows the temporal variations of the diameter of the contact area $d$ of representative Newtonian, shear-thinning and shear-thickening drops at the same $Re_0 = 200$ and $We = 556$, where the zero-shear Reynolds number $Re_0$ is defined based on the zero-shear viscosity, $Re_0 = \rho U_0 D_0 / \mu_0$.  Regardless of rheological properties, the diameter of the contact area, $d(t)$,  follows the well-established square-root temporal scaling of Newtonian drops at early times ($t<0.1$), $d = \sqrt{6t}$ (Fig.~\ref{fig:Figure 2}\textit{A} inset) \cite{Gordillo2018_impactforce,sun2022stress}. The early-time dynamics of impacting drops are primarily governed by inertia and are thus largely insensitive to surface wettability and fluid properties \cite{Yarin2006Review, mongruel2009early}. 
$d(t)$ reaches a maximum $d_{\text{max}}$ at $t = t_{\text{max}}$ before decreasing due to the recoil of the drop on the superhydrophobic surface, which ultimately leads to the bouncing of the drop at long times. Note that the diameter of the maximum contact area of the drop, $d_\text{max}$, is slightly smaller than the maximum spreading diameter, $D_\text{max}$. The latter accounts not only for the contact diameter $d_\text{max}$ but also for the thickness of the lamella rim (Fig.~\ref{fig:Figure 2}\textit{B}, inset). As expected, the shear-thinning drop spreads more, while the shear-thickening drop spreads less compared to the Newtonian drop, due to the respective reduction or increase in viscosity near the contact line.

Figure~\ref{fig:Figure 2}\textit{B} further shows the influence of various rheological parameters on the maximum spreading diameter $D_{\text{max}}$. For both shear-thinning and shear-thickening liquids, a small $\lambda = 0.01$ sets the transition from the zero-shear to infinite-shear viscosity to occur at very high shear rates. Except for very early times, the viscosity inside the drop remains nearly constant at $\mu_0$, leading to $D_{\text{max}}$ similar to that of a Newtonian drop with viscosity $\mu_0$. As $\lambda$ increases, the transition shear rate shifts toward lower values into the regime relevant to drop impact. The deviation of the maximum spreading from the Newtonian drop becomes pronounced due to increased viscosity heterogeneity within the drop. For shear-thinning liquids, the maximum spreading increases, while for shear-thickening liquids, it decreases with $\lambda$. Finally, at large $\lambda$, the viscosity within the drops quickly saturates to the infinite-shear viscosity, leading to $D_{\text{max}}$ again similar to that of a Newtonian drop but at viscosity $\mu_\infty$. The characteristic shear rate proposed in Model (\textit{i}) is valid only in this last regime \cite{laan2014maximum, de2021droplet}. As the transition to the infinite-shear plateau becomes sharper at smaller $n$, the maximum spreading of non-Newtonian drops also shows a steeper transition from the impact behavior of the Newtonian drop of viscosity $\mu_0$ to the impact behavior of the Newtonian drop of viscosity $\mu_\infty$. 

To systematically investigate the dependence of the maximum spreading of non-Newtonian drops on various impact parameters and rheological properties, we explore a vast parameter space and simulate more than 1200 drop impact events with $We$ ranging from 10 to 800, $Re_0$ from 40 to 800, $\beta$ from 0.01 and 100, $\lambda$ from 0.01 to 100, and $n$ from 0.1 to 1.

\subsection*{Energy budget}

We begin by analyzing the energy budget of drop impact, laying the foundation for deriving the characteristic shear rate of non-Newtonian drops. Energy conservation dictates 
\begin{equation}
E_{k0}=E_k+\Delta E_s+E_d+\Delta E_g, \label{eq:energy_budget}
\end{equation}
where $E_{k0} = \pi/12$ is the initial kinetic energy of the drop, $E_k = \frac{1}{2} \int_{\Omega} |\mathbf{u}|^2 \, d\Omega$ is kinetic energy, $\Delta E_s = \sigma \left(\int_{\mathcal{A}} \, d\mathcal{A} - \pi\right)$ is the change of the surface energy of the drop, and $\Delta E_g = \frac{\pi}{12}g \left(2\int_{\Omega} z\, d\Omega- 1\right)$ is the change of the gravitational energy. Here, $\sigma = 1/We$ and $g = 1/Fr^2$ are the dimensionless surface tension and gravitational acceleration, respectively, and $Fr$ is the Froude number. The integrals are calculated over the dimensionless volume $\Omega$ and surface area $\mathcal{A}$ of the drop. Within the range of the parameters explored in our study, $g < 0.08$. Thus, the effect of gravity is negligible compared to inertial effects. Lastly, the viscous dissipation at maximum spreading $t=t_{\text{max}}$, $E_d$, is given by \cite{landau1987fluid}
\begin{equation}
E_d = \int_{0}^{t_{\text{max}}} \int_{\Omega} 2 \mu ( \dot{\gamma}) \, (\mathbf{D} : \mathbf{D}) \, d\Omega \, dt. 
\label{eq:3}
\end{equation}
Here, $\mu ( \dot{\gamma})$ is the shear-rate-dependent viscosity from the Carreau model, which is non-dimensionalized by $\rho D_0 U_0$. 

Figure~\ref{fig:Figure 3}\textit{A} shows the energy conversion during the impact of a shear-thickening drop. Upon impact, the initial kinetic energy converts into excess surface energy and viscous dissipation. With an impact number $P = WeRe_0^{-2/5} \approx 66.8$ in this example, viscous dissipation dominates the energy budget at the moment of maximum spreading $t_{\text{max}}$, while the residual kinetic energy and surface energy are negligible. Figures~\ref{fig:Figure 3}\textit{B} and \textit{C} further illustrate the energy distributions at $t_{\text{max}}$ for both shear-thinning and shear-thickening drops, highlighting their dependence on $\lambda$ and $n$. With increasing $\lambda$, viscous dissipation increases for shear-thickening drops and decreases for shear-thinning drops (Fig.~\ref{fig:Figure 3}\textit{B}), as expected due to the faster transition from the zero-shear to the infinite-shear viscosity during impact. The dissipation follows an opposite trend with increasing $n$. As the transition becomes more gradual at larger $n$, viscous dissipation decreases for shear-thickening drops and increases for shear-thinning liquids (Fig.~\ref{fig:Figure 3}\textit{C}). The results are consistent with the observation on the maximum spreading diameter (Fig.~\ref{fig:Figure 2}\textit{B}): As $n$ becomes larger with a more gradual transition from $\mu_0$ to $\mu_\infty$, the effect of non-Newtonian rheology is reduced.     

 \begin{figure*}[t]
    \includegraphics[width=1\linewidth]{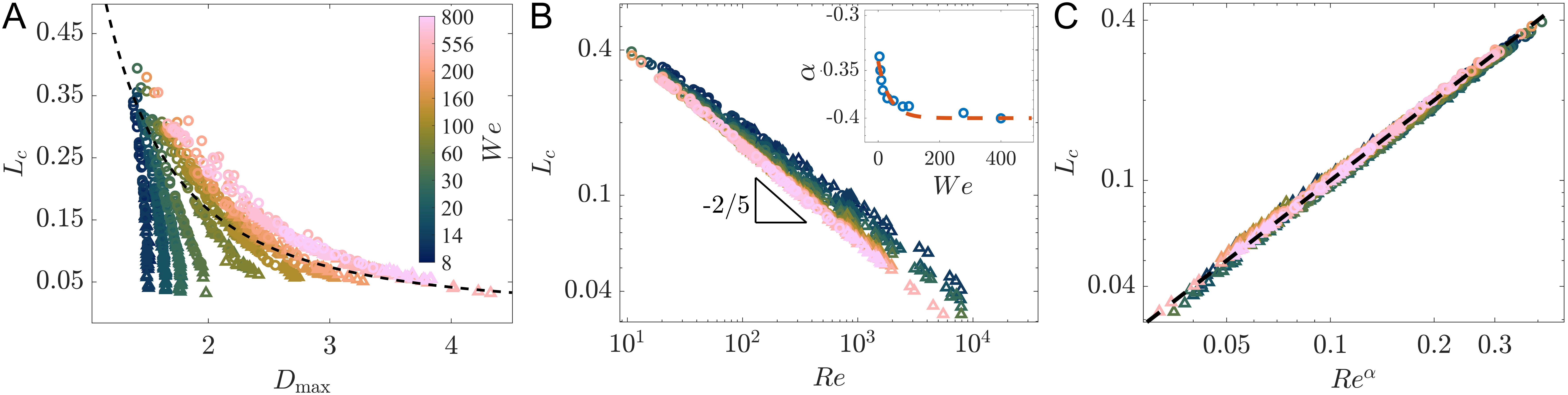}
    \caption{Characteristic length scale of non-Newtonian drop impact. (\textit{A}) The characteristic length $L_c$ versus the maximum spreading diameter $D_{\text{max}}$. For each $We$ indicated by color, $L_{c}$ follows a single curve for both shear-thinning and shear-thickening drops with $Re_0$ varying from 40 (upper left) to 800 (lower right). Triangles ($\bigtriangleup$) are for shear-thinning liquids, and circles ($\circ$) for shear-thickening liquids. The dashed line indicates the hypothesis of Model (\textit{iii}): $L_c=H_{\text{min}} = 2/(3D_{\text{max}}^2)$. (\textit{B}) $L_{c}$ versus the characteristic Reynolds number $Re$ at different $We$. The scaling $L_c \sim Re^{-2/5}$ is indicated. Inset: The power exponent $\alpha$ from $L_c \sim Re^{\alpha}$ as a function of $We$. The dashed line represents Eq.~\ref{eq:exponent}. (\textit{C}) $L_c$ versus $Re^\alpha$. The dashed line indicates the line of equality $y=x$.}
    \label{fig:Figure 4}
\end{figure*}

\subsection*{Characteristic shear rate and length}

The characteristic shear rate $\dot{\gamma}_c$ is used to capture the viscous dissipation of a non-Newtonian drop during impact \cite{boyer2016drop, shah2022coexistence, luu2009drop, liu2024universal, an2012maximum}. At $\dot{\gamma}_c$, the drop has an effective viscosity $\mu_c \equiv \mu(\dot{\gamma}_c)$, resulting in viscous dissipation at the maximum spreading that is the same as the dissipation of a Newtonian drop with constant viscosity $\mu_c$. This interpretation of $\dot{\gamma}_c$ allows us to approximate the viscous dissipation in Eq.~\ref{eq:3} as 
\begin{equation}
E_d = \mu\left( \dot{\gamma}_c\right)\dot{\gamma}_c^2 \Omega t_{\text{max}},
\label{eq:4}
\end{equation}
where the drop volume $\Omega = \pi/6$ and $\mu(\dot{\gamma})$ is from the constitutive model of the non-Newtonian liquid, which is given by the Carreau model in our simulations (Eq.~\ref{eq:Carreau_model}). In general, it can be any viscosity model chosen to describe the non-Newtonian rheology of the liquid. If we approximate the time of maximum spreading as $t_{\text{max}} = D_{\text{max}}$ \cite{chandra1991collision,wildeman2016spreading} and express the characteristic shear rate in terms of the characteristic length $\dot{\gamma}_c = 1/L_c$, Eq.~\ref{eq:4} becomes 
\begin{equation}
E_d = \frac{\pi}{6}\mu\left(\dot{\gamma}_c= \frac{1}{L_c}\right) \frac{D_{\text{max}}}{L_c^2}.
\label{eq:L_c2}
\end{equation}
Note that all quantities reported above are in dimensionless form, with velocity scaled by $U_0$ and length by $D_0$. In dimensional terms, the drop volume is $\Omega = \pi D_0^3 / 6$, the time of maximum spreading is $t_{\text{max}} = D_{\text{max}} / U_0$, and the characteristic shear rate is $\dot{\gamma}_c = U_0/L_c$.

With $E_d$ and $D_{\text{max}}$ extracted from simulations, we calculate the characteristic length $L_c$ from Eq.~\ref{eq:L_c2} and examine how $L_c$ varies with impact parameters and rheological properties. Figure~\ref{fig:Figure 4}\textit{A} shows $L_{c}$ as a function of $D_\text{max}$ for different $We$ and $Re_0$. Contrary to the proposal of Model (\textit{ii}) that the dimensionless characteristic length $L_c$ is constant at one \cite{boyer2016drop, shah2022coexistence}, our results show that $L_c$ varies significantly and is consistently less than 0.45---more than twice as small---across the range of impact parameters considered. Moreover, we also plot $H_\text{min}$ suggested by Model (\textit{iii}) (the dashed line in Fig.~\ref{fig:Figure 4}\textit{A}). At high $We$ and $Re_0$, $L_c$ appears to converge to $H_\text{min}$. However, at lower $We$ and $Re_0$, $L_c$ deviates substantially from $H_\text{min}$, contradicting the model hypothesis. 

Importantly, at a given $We$, $L_c$ follows a single curve and shows a one-to-one correspondence with $D_{\text{max}}$ regardless of rheological parameters or even whether the drop exhibits shear thinning or shear thickening (Fig.~\ref{fig:Figure 4}\textit{A}). If the maximum spreading of a non-Newtonian drop can be mapped to that of a Newtonian drop with viscosity $\mu_c$, $D_{\text{max}}$ should only be a function of $Re$ and $We$, i.e., $D_{\text{max}}=D_{\text{max}}(Re,We)$, where, different from the zero-shear Reynolds number $Re_0$, $Re$ is the characteristic Reynolds number defined as $Re = \rho U_0 D_0/\mu_c$. The complex shear-rate-dependent viscosity of the non-Newtonian liquid is captured by a single parameter, $\mu_c = \mu(\dot{\gamma}_c) = \mu(1/L_c)$, via $Re$. Thus, if the mapping exists, the dependence of $L_c$ on $D_{\text{max}}$ shown in Fig.~\ref{fig:Figure 4}\textit{A} requires that $L_c$ must also be a function of $Re$ and $We$ alone, $L_c = L_c\left[D_{\text{max}}(Re,We),We\right] = L_c(Re,We)$, independent of any other features of the liquid's constitutive relation.         

To verify the existence of the non-Newtonian to Newtonian mapping, we plot $L_c$ as a function of $Re$, which nearly collapses all our data with a scaling $L_c \sim Re^{-2/5}$ (Fig.~\ref{fig:Figure 4}\textit{B}). To account for the small $We$ correction, we fit the power-law exponent $\alpha$ in $L_c \sim Re^{\alpha}$ at different $We$. $\alpha$ approaches $-0.4$ at high $We$ and decreases slightly to about $-0.34$ at the lowest $We$ of our simulation (Fig.~\ref{fig:Figure 4}\textit{B} inset). We approximate $\alpha(We)$ as 
\begin{equation} 
\alpha(We) = -0.4 + 0.06e^{-0.03We}, \label{eq:exponent}    
\end{equation}
which allows us to collapse all our data onto a master curve:
\begin{equation}
    L_c = CRe^{\alpha(We)} \label{eq:L_c} 
\end{equation} 
with an order-one prefactor $C=0.986 \pm 0.002$ obtained from fitting (Fig.~\ref{fig:Figure 4}\textit{C}). Equation~\ref{eq:L_c} confirms that the characteristic length depends solely on $Re$ and $We$ and therefore demonstrates the existence of a characteristic shear rate that allows for the mapping from the maximum spreading of non-Newtonian drops to that of Newtonian drops. This finding justifies the hypothesis that has been widely assumed without validation \cite{laan2014maximum, de2021droplet,boyer2016drop, shah2022coexistence,luu2009drop,liu2024universal, an2012maximum,quirke2024spreading}. Given a non-Newtonian constitutive relation, Eq.~\ref{eq:L_c} can be solved to obtain the characteristic length and shear rate, enabling the calculation of the characteristic Reynolds number of non-Newtonian drop impact. Note that Eq.~\ref{eq:L_c} is nonlinear, requiring a numerical solution due to the dependence of $Re$ on $L_c$ through the constitutive model.

\subsection*{Origin of the characteristic length}

The characteristic length $L_c$ depends primarily on $Re$ following $L_c \approx Re^{-2/5}$ (Figs.~\ref{fig:Figure 4}\textit{B} and \textit{C}). Such a relation can be understood from the asymptotic film thickness of the spreading drop \cite{roisman2009inertia, eggers2010drop, lagubeau2012spreading, Gordillo2018_impactforce}. In the inertia-dominant regime at high $Re$ and $We$, the flow within a spreading drop is approximately hyperbolic with $v_r = r/(t+t_0)$ and $v_z = -2z/(t+t_0)$, where $v_r$ and $v_z$ are the radial and vertical velocity, respectively. The time $t_0$ marks the onset of the inertia-driven spreading following the early-impact dynamics. The evolution of the drop's free surface, denoted as $h(r,t)$, is governed by the equation of motion, 
\begin{equation}
    \frac{\partial h}{\partial t} + v_r \frac{\partial h}{\partial r} = v_z.
\end{equation} 
A self-similar solution to this equation is given as \cite{eggers2010drop}:
\begin{equation}
h(r,t)=\frac{1}{\left( t+t_0\right)^2} \mathcal{H}\left(\frac{r}{t+t_0}\right). \label{eq:self-similar}    
\end{equation}
Equation~\ref{eq:self-similar} predicts the height of the drop as $h(0,t) = \mathcal{H}(0)/\left( t + t_0 \right)^2$, which simplifies to $h(0,t) = \mathcal{H}(0)/t^2$ when $t \gg t_0$ in the asymptotic limit. Previous studies have investigated drop height, yielding $\mathcal{H}(0) = 0.39$ and $t_0 = 0.25$ from simulations \cite{roisman2009inertia}, $\mathcal{H}(0) = 0.49$ and $t_0 = 0.43$ from experiments \cite{lagubeau2012spreading}, and $\mathcal{H}(0) = 0.44$ and $t_0 = 0.31$ from theory \cite{Gordillo2018_impactforce}. 

While the free surface of the drop descends, a boundary layer grows upward from the impacted surface, following $h_{\nu} = \sqrt{t/Re}$. The boundary layer intersects the free surface at $t^* = \mathcal{H}(0)^{2/5} Re^{1/5}$, halting the spreading \cite{Schroll2010_Viscousspreading}. Therefore, the smallest film thickness in the asymptotic limit is $h(0, t^*) =  \mathcal{H}(0)^{1/5} Re^{-2/5}$ with the prefactor $\mathcal{H}(0)^{1/5} = 0.85 \pm 0.02$ on the order of one. Identifying this smallest drop height as the characteristic length $L_c$, we recover the relationship between $L_c$ and $Re$. At low $We$, surface tension interferes with the inertia-driven spreading, altering the shape of the drop and likely contributing to the weak $We$ correction given in Eq.~\ref{eq:exponent}. 

Thus, the assumption of Model (\textit{iii})---that the maximum spreading is governed by the strongest shear at the smallest drop height---remains valid, even though the estimate of the smallest height in Model (\textit{iii}) is incorrect. Here, we show that the characteristic length corresponds to the smallest drop height, determined by the interplay between the free surface descent and boundary layer growth.   

\begin{figure*}[t]
    \centering
    \includegraphics[width=1\linewidth]{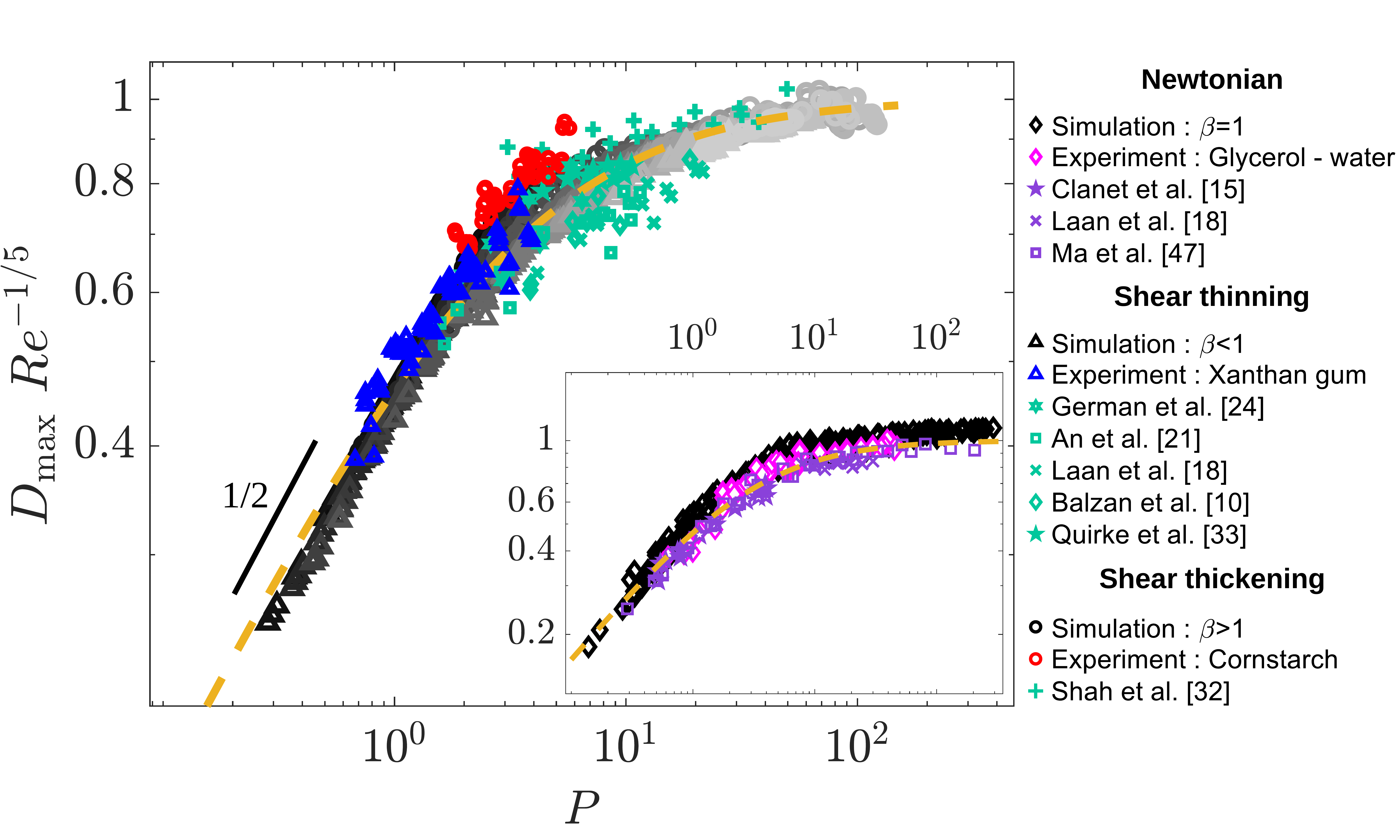}
    \caption{Maximum spreading of Newtonian and non-Newtonian drops. The rescaled maximum spreading diameter, $D_{\text{max}}Re^{-1/5}$ is plotted against the impact number $P = We^{*}Re^{-2/5}$. Black and gray symbols represent data from our simulations, while colored symbols correspond to our experiments and experimental results from the literature, as detailed in the figure legend. The yellow dashed line is a second-order Pad\'e approximant that bridges the viscous regime at large $P$ and the capillary regime at low $P$ (Eq.~\ref{eq:PadeApp}). The scaling $D_\text{max}Re^{-1/5} \sim P^{1/2}$ in the capillary regime is indicated by the solid line. To reduce clutter, results for Newtonian drops are shown separately in the inset, where the same Pad\'e approximant is included for comparison. Experiments on Newtonian drops are performed using glycerol-water mixtures with varying glycerol/water ratios (SI Sec.~C). The results confirm that the maximum spreading of non-Newtonian drops can be mapped onto that of Newtonian drops using the characteristic $Re$ at $\dot{\gamma}_c=1/L_c$ determined by Eq.~\ref{eq:L_c}.}
    \label{fig:Figure 5}
\end{figure*}

\subsection*{Maximum spreading of non-Newtonian drops} \label{exp_section}
 
With the characteristic shear rate and therefore the characteristic Reynolds number established via Eq.~\ref{eq:L_c}, we can now map the maximum spreading of non-Newtonian drops to that of Newtonian drops. Figure~\ref{fig:Figure 5} shows the rescaled maximum spreading diameters $D_{\text{max}} Re^{-1/5}$ of both Newtonian and non-Newtonian drops as a function of the impact number $P = We^{*}Re^{-2/5}$ \cite{eggers2010drop, Yufei2024Capillary}. Here, the modified Weber number is defined as $We^{*}=(We+12)/(1-\cos\theta)$ to account for the effect of surface wettibility on the drop spreading at low $We$ \cite{lee2016universal,Yufei2024Capillary}. The dynamic contact angle, $\theta$, can be evaluated from the static contact angle and the capillary number (SI Sec.~B). The expression of $We^*$ is derived from an energy balance at low $We$ (SI Sec.~B). 

Remarkably, all our numerical data converge onto a master curve, irrespective of the rheological properties of the drop (Fig.~\ref{fig:Figure 5}). The viscous regime is reached at $P \gg 1$ with $D_{\text{max}} \sim Re^{1/5}$. At small $P$ and large $We$, $D_{\text{max}} \sim \left[(We + 12)/(1 - \cos{\theta})\right]^{1/2} \sim We^{1/2}$, confirming the scaling relation in the capillary regime \cite{laan2014maximum, eggers2010drop, Yufei2024Capillary}. In the limit of small $We$, $D_\text{max}$ asymptotes to a constant set by the contact angle $\theta$ \cite{lee2016universal,Yufei2024Capillary,roisman2009inertia}. We bridge the viscous and capillary regimes using a second order Pad\'e approximant \cite{wang2022maximal},
\begin{equation}
    D_{\text{max}} Re^{-1/5} = \frac{P^{1/2} + A P}{B + P^{1/2} + A P} \label{eq:PadeApp}
\end{equation} 
with two constants $A = 0.76 \pm 0.07$ and $B=2.0 \pm 0.1$ from fitting. The fit yields $D_\text{max} \approx (0.50\pm0.03)We^{*1/2}$ in the capillary regime with $P\ll1$, in good agreement with the energy balance with negligible viscous dissipation (SI Sec.~B).

Finally, to verify our numerical prediction, we conduct experiments using aqueous solutions of xanthan gum as a model shear-thinning liquid \cite{whitcomb1978rheology,xuewu1996rheological} and cornstarch-water mixtures as a model shear-thickening liquid \cite{Brown2009_cornstarch,Fall2012_cornstarch}. The mass fraction of xanthan gum is varied from 0.01 wt\% to 0.5 wt\%, while the mass fraction of cornstarch is adjusted between 10 wt\% and 40 wt\% in the continuous shear-thickening regime \cite{Brown2014_shearthickening}. The rheological properties of these samples are detailed in SI Sec.~C and Table S1. We fix the drop diameter at $D_0 =3.4$ mm and vary the impact velocity $U_0$ between 0.5 and 1.5 m/s. We use glass slides coated with a superhydrophobic spray (SOFT99 Glaco Mirror Coat Zero) as the impacted surfaces, which exhibit a static contact angle of $\theta_0 = 150 \pm 5^\circ$ and a contact angle hysteresis of less than $14^\circ$. Side-view images captured at 40,000 frames per second using high-speed photography are analyzed to extract $D_{\text{max}}$ (Movies 1 and 2).

In addition to our own measurements, we incorporate data from prior experimental studies on non-Newtonian drop impact. These studies were conducted on impacted surfaces having a wide range of wettability and employed various constitutive models to describe the rheology of different non-Newtonian liquids, including the Cross model for aqueous solutions of xanthan gum and colloidal suspensions of silica particles \cite{an2012maximum,shah2022coexistence}, the power-law fluid model for xanthan gum solutions and blood \cite{laan2014maximum, german2009impact}, and the Herschel–Bulkley model for graphene oxide suspensions and dairy-based solutions \cite{balzan2021drop, quirke2024spreading}.  

Experimental results from both our measurements and the literature exhibit excellent agreement with our numerical predictions (Fig.~\ref{fig:Figure 5}). Thus, our study combining simulations, scaling analyses and experiments successfully maps the spreading dynamics of non-Newtonian drops to those of Newtonian drops, providing a robust quantitative description of the maximum spreading of generalized Newtonian liquids regardless of specific constitutive models.

\section*{Conclusion and outlook}

Despite the pervasive presence of non-Newtonian drops in natural and industrial contexts, the impact dynamics of such drops remain elusive. Through a combination of simulations, scaling analyses, and experiments, our study addresses this important knowledge gap and sheds light on the maximum spreading---one of the most important features of drop impact---of impacting non-Newtonian drops. 

Fundamentally, our study provides the long-overdue validation of the existence of a characteristic shear rate that can effectively capture the influence of non-Newtonian rheology on the maximum spreading of drop impact. We demonstrate that this characteristic shear rate is governed by a length scale corresponding to the asymptotic film thickness of a spreading drop, defined at the moment when the descending free surface of the drop intersects with the boundary layer growing from the substrate. 

Practically, we offer a simple recipe for calculating the maximum spreading diameter of non-Newtonian drops. Starting with any given constitutive model that describes the rheology of a generalized Newtonian liquid, one can determine the characteristic length and Reynolds number using Eq.~\ref{eq:L_c}. Once the characteristic Reynolds number is obtained, the maximum spreading diameter of the non-Newtonian drop can be computed using the established relations for the maximum spreading diameter of Newtonian drops, such as the Pad\'e approximant employed in this study or any other models appropriate for the specific impact regime \cite{Aksoy2022Spreaing}. This procedure is reversible. For a desired maximum spreading diameter, one can reverse the calculation to identify the rheological properties of liquid required to achieve it. Consequently, beyond advancing the fundamental understanding of non-Newtonian fluid dynamics in the classical problem of drop impact, this work provides an actionable guideline for tailoring the rheology of non-Newtonian liquids---such as paints, inks, and food formulations---to enable precise control over the outcome of drop impact.

While our study focuses on generalized Newtonian liquids that exhibit shear thinning and shear thickening behaviors, it is important to recognize the broader spectrum of non-Newtonian fluids, each characterized by distinct rheological properties beyond the scope of this study \cite{Macosko1994}. In particular, extensive studies have investigated the impact dynamics of viscoplastic drops, where the initial kinetic energy is dissipated through plastic deformation caused by the high yield stress of the fluids \cite{luu2009drop, luu2013giant, blackwell2015sticking, oishi2019normal, isukwem2024role, isukwem2024viscoplastic, isukwem2025viscoplastic}. Viscoelasticity is another key rheological feature of polymeric fluids that influences drop impact \cite{Izbassarov_2016_viscoelasticity,Wang_2017_viscoelasticity,venkatesan2019computational}. For the most concentrated xanthan gum solution in our study (0.5 wt\%), the Weissenberg number reaches values of order one at the highest impact velocity explored, indicating the presence of weak elasticity during drop impact. Exploring how finite elasticity, in conjunction with shear-thinning rheology, affects maximum spreading is an intriguing avenue for future research. Additionally, shear-thickening liquids, such as cornstarch-water mixtures at high mass fractions, can undergo a liquid-to-solid transition upon impact via discontinuous shear thickening and shear jamming \cite{Brown2014_shearthickening, shah2022coexistence}. This transition is not captured by the current study, which focuses on continuous shear thickening at relatively low cornstarch mass fractions. The general framework developed in this study can be extended to analyze this broader range of non-Newtonian liquids by incorporating elastic energy and plastic dissipation \cite{isukwem2024role} into the energy budget. 

\subsection*{Data, Materials, and Software Availability} All study data are included in the article and Supporting Information.

\acknow{We thank Vatsal Sanjay for assistance with Basilisk simulations, Michelle Driscoll, and Brian Seper for fruitful discussions, and Matthias M\"{o}bius, Leonardo Gordillo and Michelle Driscoll for the comments on our manuscript. We also thank Michelle Driscoll and Phalguni Shah for providing data on colloidal suspensions, and Matthias M\"{o}bius and Jennifer Quirke for sharing data on graphene-oxide suspensions. We thank the Minnesota Supercomputing Institute (MSI) at the University of Minnesota for computational resources. This research was supported by NSF DMR-2002817. A.M. acknowledges the financial support from the PPG Fellowship.}

\showacknow{} 

\bibsplit[4]

\bibliography{PNAS_bibliography}
\end{document}